\documentclass[useAMS,usenatbib]{mn2e}

\usepackage{graphicx}

\title{An estimate of the SN kick velocities for High Mass X-ray
Binaries in the SMC}

\author[M.J. Coe]{M. J.~Coe  \\
School of Physics and Astronomy, Southampton University, SO17 
1BJ, UK\\}
\begin{document}

\date{Jan 2005}

\pagerange{\pageref{firstpage}--\pageref{lastpage}} \pubyear{2002}

\maketitle

\label{firstpage}

\begin{abstract}

This work investigates the possible supernova kick velocities imposed on HMXB
systems in the Small Magellanic Cloud. Comparisons are made between
the location of such systems and the locations of young, stellar
clusters on the premise that these may represent the birthplace of
many of these systems. Measurements of the separation of clusters and
HMXBs,and an estimate of the typical lifetimes of these systems, 
leads to a minimum average space velocity 30 km/s. This value
is compared to theoretical estimates.

\end{abstract}

\begin{keywords}
stars:neutron - X-rays:binaries - Magellanic Clouds
\end{keywords}

\section{Introduction and background}

The Be/X-ray systems represent the largest sub-class of massive X-ray
binaries.  A survey of the literature reveals that of the 115
identified massive X-ray binary pulsar systems (identified here means
exhibiting a coherent X-ray pulse period), most of the systems fall
within this Be counterpart class of binary.  The orbit of the Be star
and the compact object, presumably a neutron star, is generally wide
and eccentric.  X-ray outbursts are normally associated with the
passage of the neutron star close to the circumstellar disk (Okazaki
\& Negueruela, 2001). A recent review of these systems may be found in
Coe (2000).

X-ray satellite observations have revealed that the Small Magellanic
Cloud (SMC) contains an
unexpectedly large number of High Mass X-ray Binaries (HMXB). At the
time of writing, 47 known or probable sources of this type have been
identified in the SMC and they continue to be discovered at a rate of
about 2-3 per year, although only a small fraction of these are active
at any one time because of their transient nature.  Unusually
(compared to the Milky Way and the LMC) all the X-ray binaries so far
discovered in the SMC are HMXBs, and equally strangely, only one
of the objects is a supergiant system, all the rest are Be/X-ray
binaries.

\section{Birthplaces and kick velocities}

There is considerable interest in the evolutionary path of High Mass
X-ray binary systems (HMXBs), and, in particular, the proper motion of
these systems arising from the kick velocity imparted when the neutron
star was created. Portegies Zwart (1995) and Van Bever \& Vanbeveren
(1997) investigate the evolutionary paths such systems might take and
invoke kick velocities of the order 100 - 400 km/s. In an
investigation into bow shocks around galactic HMXBs Huthoff \& Kaper
(2002). They then used Hipparcos proper motion data used to derive
associated space velocities for Be/X-ray and supergiant systems.  From
their results, an average value of 48 km/s is found for the 7 systems
that they were able to fully determine the three dimensional
motion. Since this is rather lower than the theoretical values it is
important to seek other empirical determinations of this motion.

This paper addresses what may be learnt about kick velocities by
looking at the possible association of HMXBs in the SMC with the nearby young
star clusters from which they may have emerged as runaway systems. 
A review of the
properties of the optically (and/or IR) identified counterparts to the
X-ray pulsars is presented in Coe et al. (2005). In that paper the
authors introduce the nomenclature of SXPnnn for the systems, where
SXP stands for Small Magellanic Cloud X-ray Pulsar, and the numbers
nnn identify the X-ray pulse period in seconds. This simplified
identification is used here. The SMC cluster data come from an
extensive review of the spatial location of stars by Rafelski
\& Zaritsky (2004). Clusters from this paper are hereafter referred to
as RZ clusters.

\section{Analysis}

It is critical to use the exact position for the SXP sources if
distances are to be measured - some have X-ray positional errors of a few
arcminutes. Consequently, only the SXP objects from Coe et al. (2005)
which had precise optical counterparts were selected for this
work. This resulted in an SXP list of 17 objects.

\begin{figure}\begin{center}
\includegraphics[width=65mm,angle=-90]{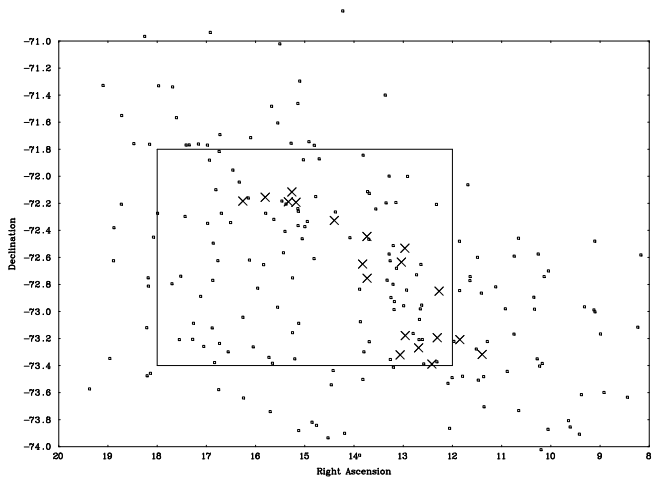}
\caption{Distribution of SXPs (shown as crosses) and RZ clusters
  (shown as squares) in the SMC. The rectangular box indicates the region used
for the random sampling comparison - see text.}
\label{fig:map}
\end{center}
\end{figure}

Firstly the spatial distributions of SXPs and the RZ clusters within
the SMC were checked to ensure adequate matching coverage around the
regions occupied by the SXP objects. This was found to be satisfactory
- see Figure~\ref{fig:map}. The OGLE catalogue of SMC clusters
(Pietrzynski et al., 1998) was also considered, but this has restricted
coverage and several of the SXP sources fell outside the OGLE region.

In order to determine whether the SXP sources may have originated from
a nearby stellar cluster, the coordinates of the SXP objects
were compared to those of the RZ clusters. For every SXP its
position was compared to the location of all of the  RZ clusters and the
identification of the nearest cluster neighbour obtained.  The 
results from this search are presented in Table~\ref{tab:sxprz}.

\begin{table}
\begin{center}

\begin{tabular}{ccc}
\hline
SXP ID&Nearest RZ Cluster&Separation\\
&ID& (arcmin)\\
\hline
SXP0.92 &  046 & 3.18 \\
SXP3.34 &  150 & 2.43 \\
SXP8.02 &  124 & 2.87 \\
SXP8.80 &   77 & 4.45 \\
SXP9.13 &   60 & 5.33 \\
SXP15.3 &   80 & 3.99 \\
SXP59.0 &   90 & 1.40 \\
SXP74.7 &   51 & 7.38 \\
SXP82.4 &   72 & 3.25 \\
SXP172  &   68 & 2.92 \\
SXP304  &  133 & 5.47 \\
SXP323  &   66 & 3.67 \\
SXP348  &  150 & 6.37 \\
SXP452  &  130 & 1.33 \\
SXP504  &   96 & 5.56 \\
SXP565  &  101 & 3.79 \\
SXP756  &   57 & 1.98 \\
\hline
\end{tabular}

\caption{Table of optically identified SXP objects and, in each case,
  the nearest RZ cluster and its distance away from the SXP source. }

\label{tab:sxprz}
\end{center}
\end{table}

\begin{figure}\begin{center}
\includegraphics[width=65mm,angle=-0]{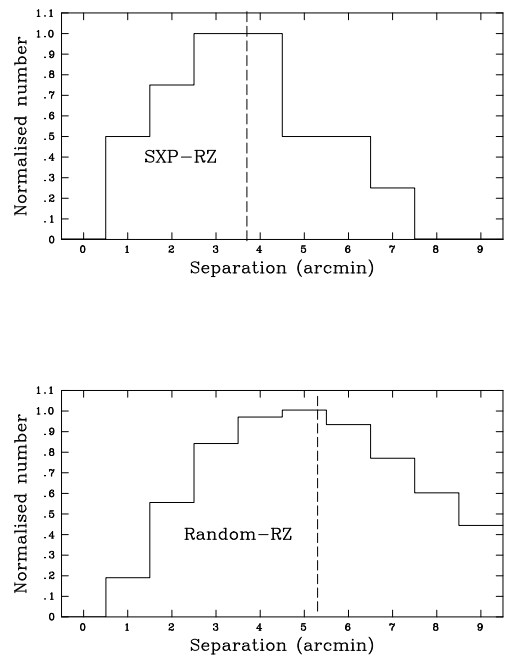}
\caption{Upper panel : histogram of the distances between each SXP and its nearest
neighbour RZ cluster. Lower panel : histogram of the minimum distance between random points and
  RZ clusters. In each case the dotted vertical line shows the
position of the mean of the distribution.}
\label{fig:both}
\end{center}
\end{figure}

The average distance between the pairs of objects listed in 
Table~\ref{tab:sxprz} was found to be 3.85 arcminutes. 
The histogram of the distances between each SXP source and the nearest
RZ cluster is shown in the upper panel of Figure~\ref{fig:both}.

Obviously it is important to ensure that the SXP-RZ cluster distances
are significantly closer than a sample of randomly distributed
points. One way to determine this is simply to just use the RZ cluster
data and find the average cluster-cluster separation. This gives a
value of 6.13 arcminutes. Alternatively, the rectangular region
indicated in Figure~\ref{fig:map} was used, and the minimum distance
between 100,000 random points and, in each case, the nearest RZ
cluster was found. The average value was found to be 5.30 arcminutes
and the corresponding histogram is shown in the lower panel of
Figure~\ref{fig:both}. From comparing the two histograms it is clear
that there does exist a much closer connection between SXP sources and
RZ clusters than expected randomly.

This conclusion may be tested further using either the
Kolmogorov-Smirnov (K-S) test or the Student t-test to quantify the
probability that the two distributions are distinctly different. The
K-S test gives a probability of 9\% that the two distributions are the
same, whereas the t-test gives a value of 3\% that the means are the
same. Thus the above conclusion (that the SXP sources are closer to
the clusters than random) is well supported, but not
overwhelmingly so.

\begin{figure*}
\begin{center}
\includegraphics[]{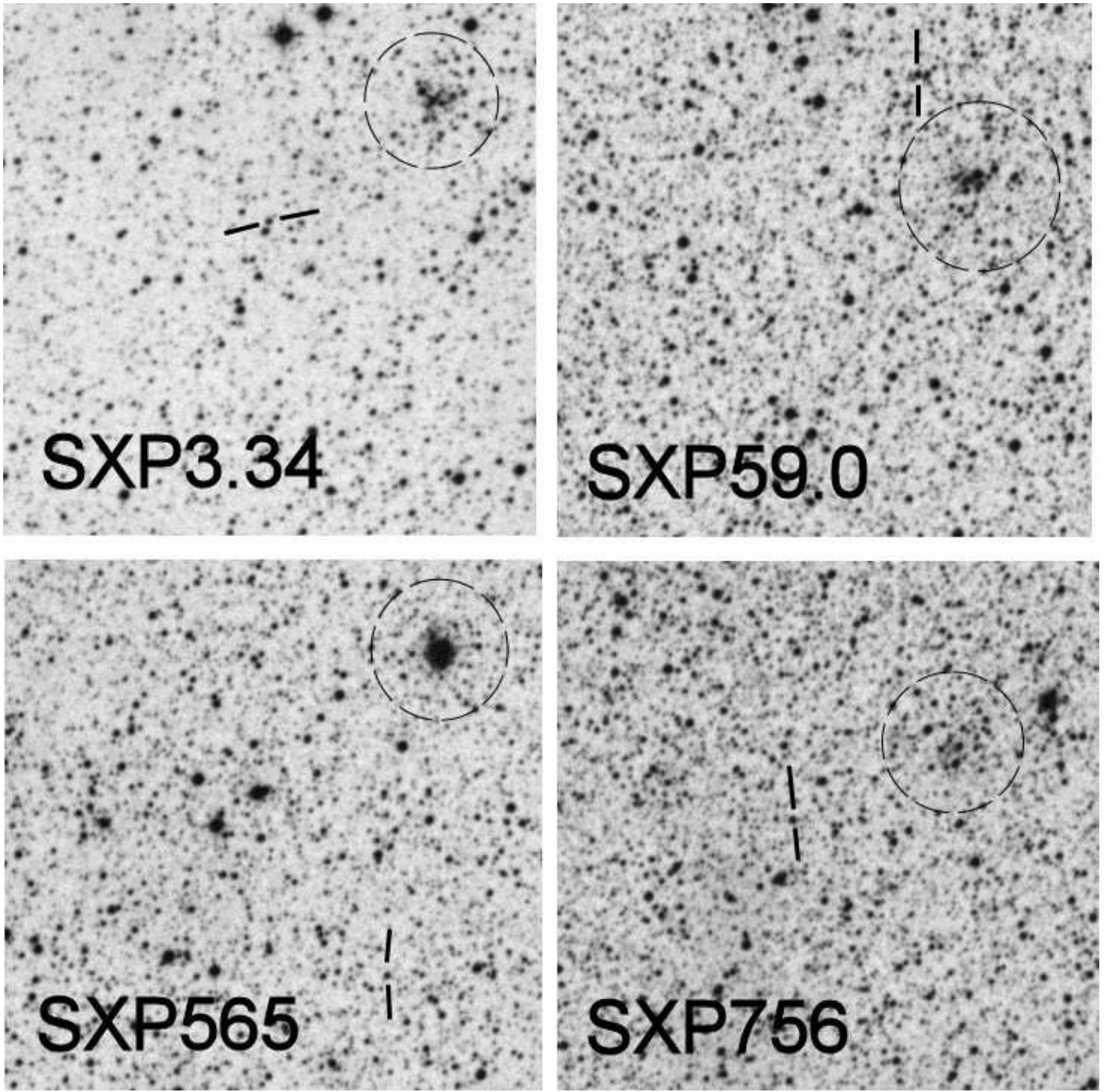}
\caption{Four examples of the fields around SXP objects showing the
  nearest RZ cluster. Each field is 6 x 6 arcminutes in size with North
upwards and East to the left.  The SXP
object is indicated and the associated RZ cluster is shown within the
dashed circle.}
\label{fig:all4}
\end{center}
\end{figure*}

A sample of 4 SXP objects and their nearest RZ cluster is shown in
Figure~\ref{fig:all4}. The image shown is from the digitised sky survey red
band.

\section{Discussion}

If the distance of 3.85 arcminutes is indicative of the distance the
SXP pulsar systems have travelled on average since birth, then this
value may be used to estimate the kick velocity of the systems. To do
this one first needs to convert the angular separation into a real
distance. Using a value of 60 kpc for the distance to the SMC, then
3.85 arcminutes corresponds to 65 pc. To compute a velocity one has to
have some estimate of the likely travel time of the SXP system since
birth.  An upper limit on this may be taken from the evolutionary
models of Savonije \& van den Heuvel (1977) who estimate the maximum
possible lifetime of the companion Be star after the creation of the
neutron star to be 5 million years. However, in the same paper they
estimate the spin down time of the pulsar since creation to be less
than 600,000 years. If we take the lower limit (the spin down time)
then the {\it minimum} average velocity of the SXP systems projected
on to the sky is 130 km/s. This number is significantly higher than
the numbers presented by other authors, for example van den Heuvel et
al. (2000). Those authors interpreted the Hipparcos results for
galactic HMXBs (Chevalier \& Ilovaisky, 1998) in terms of models for
kick velocities, and concluded that values around 15 km/s were more
appropriate for such systems. So if, instead, the lifetime number
of 5 million years is used, then transverse velocity components of 16
km/s are obtained.

In either case, assuming the systems are moving in completely random
directions compared to our line of sight, then the true average space
velocity will be a factor $\sim$2 greater, i.e. an average of 32 km/s
in the second case.

Finally, the ages of the RZ clusters associated with the SXP sources
are of significant interest. Using the extinction corrected ages
presented in Table 2 of Rafelski \& Zaritsky (2004) it is possible to
determine the mean age to be 130$\pm$140 Myrs. As can be seen from the
histograms presented in their paper, this is very much at the young
end of their sample distribution. It therefore reinforces the
suggestion that the clusters identified with the SXP sources are very
likely to be the correct parent clusters for these objects.

\section{Conclusions}

As a result of the study of HMXBs in the SMC it has been shown that
there exists a convincing link between the these systems and nearby
stellar clusters. Accepting this link, leads to a determination
of the average space velocity of the systems arising from the SN kick
to be $\sim$30 km/s. This value is in good agreement with
observational and 
theoretical values.

\bsp

\label{lastpage}


\begin{thebibliography}{99}

\bibitem []{} Chevalier C., Ilovaisky S.A., 1998, A\&A 330, 201.

\bibitem[]{} Coe M.J. 2000 in ``The Be Phenomenon in Early-Type
Stars'', IAU Colloquium 175, ASP Conference Proceedings, Vol. 214, ed:
M. A. Smith H. F. Henrichs.  Astronomical Society of the Pacific,
p.656.


\bibitem[]{} Coe M.J., Edge W.R.T.E., Galache J.L., McBride V.A.,
  2005, MNRAS (in press).

\bibitem[]{} Huthoff F. \& Kaper L., 2002, A\&A 383, 999.


\bibitem[]{} Okazaki A.T. \& Negueruela I., 2001 A\&A 377, 161.

\bibitem[]{} Pietrzynski, G., Udalski, A., Kubiak, M., Szymanski, M.,
Wozniak, P., Zebrun, K., 1998 AcA 48, 175.

\bibitem[]{} Portegies Zwart S.F., 1995, A\&A 296, 691.

\bibitem[]{} Rafelski M. \& Zaritsky D., 2004, AJ (in press).

\bibitem[]{} Savonije G.J. \& van den Heuvel E.P.J., 1997, ApJ 214, L19.

\bibitem[]{} van den Heuvel E.P.L., Portegies Zwart S.F., Bhattacharya
D., Kaper L., 2000, A\&A 364, 563.

\bibitem[]{} Van Bever J., Vanbeveren D., 1997, A\&A 322, 116.



\end{thebibliography}
\end{document}